\documentclass[journal]{IEEEtran}

\usepackage{amsmath,amssymb}
\usepackage{graphicx}

\usepackage{booktabs}
\usepackage{hyperref}
\usepackage{xcolor}
\usepackage{algorithm}
\usepackage{algpseudocode}
\usepackage{multirow}
\usepackage{siunitx}
\usepackage{cite}
\usepackage{pifont}
\newcommand{\cmark}{\ding{51}}%
\newcommand{\xmark}{\ding{55}}%

\begin{document}

\title{X-LIBS: Interpretable Soil Classification Using Explainable AI and
       Laser-Induced Breakdown Spectroscopy}

\author{Yingchao~Huang,~\IEEEmembership{}
        Xin~Wang%
\thanks{Y. Huang and X. Wang are with the Faculty of Digital Innovation, Arts
and Sciences, Saskatchewan Polytechnic, Regina, SK, Canada (e-mail:
huangyi@saskpolytech.ca; wangxi@saskpolytech.ca).}}

\markboth{IEEE Transactions on Plasma Science}{}

\maketitle

\begin{abstract}
Machine learning (ML) has emerged as a powerful tool for soil analysis using
Laser-Induced Breakdown Spectroscopy (LIBS). However, traditional black-box models
often lack interpretability, limiting their effectiveness in decision-making processes.
This study introduced explainable AI (XAI) techniques to enhance both the
interpretability and classification performance of Partial Least Squares Discriminant
Analysis (PLS-DA) models for soil classification with LIBS. Local Interpretable
Model-agnostic Explanations (LIME) was employed to
identify the local spectral features associated with soil elements, providing
explanations for their contributions to classification outcomes. To improve
classification performance on unlabeled spectra, prediction uncertainty was quantified
by sequentially removing the top local spectral features identified by LIME as critical
to the classification. Higher uncertainty was associated with a smaller number of feature
removals needed to change the label. Spectra for which a large number of features had
to be removed before the label changed were treated as high-confidence predictions and
admitted to a co-training process. Thresholds on this flip count were defined separately for
spectrally ambiguous and stable classes to manage uncertainty identification effectively.
To address the sensitivity of PLS-DA to class imbalances, equal proportions of
pseudo-labels from each class were included in co-training. The proposed method was
evaluated on the publicly available EMSLIBS dataset, achieving a test accuracy of
92.69\%, competitive with the best-performing methods in the literature. The XAI results
provided insights into the causes of misclassifications and identified the dominant
spectral features critical for accurate predictions, advancing interpretable AI solutions
for LIBS-based soil analysis.
\end{abstract}

\begin{IEEEkeywords}
Laser-induced breakdown spectroscopy, explainable artificial intelligence, LIME,
soil classification, PLS-DA, semi-supervised learning.
\end{IEEEkeywords}

\section{Introduction}
\label{sec:intro}

\IEEEPARstart{L}{aser}-Induced Breakdown Spectroscopy (LIBS) is a rapid, micro-destructive atomic
emission technique in which a focused, high-irradiance laser pulse ablates a microgram-scale quantity
of solid, liquid, or gaseous material to generate a short-lived luminous
plasma~\cite{cremers2013libs,noll2012book,miziolek2006book}. As the plasma cools over a period of
microseconds to milliseconds, electronically excited atoms, ions, and molecules undergo radiative
transitions whose wavelengths and intensities encode the full multi-elemental composition of the ablated
target~\cite{hahn2010part1,hahn2012part2}. Spectrally resolved collection of this emission yields, in a
single laser shot and without sample preparation, chemical reagents, or vacuum conditions, a fingerprint
spanning the ultraviolet-to-near-infrared range that simultaneously covers virtually every element in the
periodic table~\cite{fortes2013review}. These attributes such as simultaneous multi-element detection, speed,
stand-off capability, and freedom from sample preparation have driven sustained interest in LIBS across
geoscience, environmental monitoring, planetary exploration, and precision
agriculture~\cite{harmon2019libs_review,radziemski2002history}.

The application of LIBS to soil and geomaterial characterisation stretches back more than three decades.
Among the earliest demonstrations, Capitelli et al.~\cite{capitelli2002heavy} established that LIBS could detect heavy metals such
as lead, copper, and chromium in contaminated soils at environmentally relevant concentrations without
wet-chemical digestion. Subsequent work broadened the scope to agronomically
critical macronutrients, establishing quantitative LIBS procedures for soil organic carbon, nitrogen, and
phosphorus~\cite{senesi2016carbon,nicolodelli2019trac}. Comprehensive two-part reviews by Villas-Boas et
al.\ surveyed LIBS applications across the full spectrum of soil physical and chemical properties,
covering texture, pH, humification degree, and elemental analysis under both laboratory and field
conditions~\cite{villasboasI2020,villasboasII2020}. Parallel advances in miniaturised and handheld LIBS
instrumentation have since enabled in-situ, in-field geochemical surveys that support soil mapping at
throughputs and spatial densities inaccessible to traditional laboratory
analysis~\cite{harmon2019libs_review,el2020review}.

LIBS spectra are intrinsically high-dimensional: a single acquisition may capture tens of thousands of
wavelength channels encoding overlapping emission lines from dozens of elements superimposed on continuum
background. For soil samples in particular, matrix effects arising from variable mineralogy, moisture
content, porosity, and grain size introduce non-linear perturbations in emission intensities that cannot
be corrected by univariate calibration alone~\cite{hahn2010part1,lucena2011libs_soil}. The spectral
complexity of mixed-composition samples further complicates classification, especially when multiple source
materials contribute overlapping emission signatures~\cite{kepes2020benchmark}. A dedicated benchmark
dataset, EMSLIBS, was created specifically to challenge classification algorithms under these realistic
conditions. These challenges have provided strong motivation for adopting
multivariate chemometric and machine learning (ML) approaches that jointly exploit all wavelength channels
to extract robust analytical information.

Among classical multivariate methods, principal component analysis (PCA) and partial least squares (PLS)
regression and discriminant analysis have been the workhorses of LIBS data modelling. Pořízka et al.~\cite{porizka2018pca}
reviewed PCA applications across the LIBS literature, documenting its roles in spectral compression,
outlier detection, and exploratory classification. Clegg et al. ~\cite{clegg2009pls} demonstrated that
PLS-based multivariate regression substantially outperforms univariate line-intensity calibration for
remote LIBS geological analysis, while Dyar et al.~\cite{dyar2012pls} extended the comparison to
include LASSO, elastic net, and support vector regression. Boucher et al.~\cite{boucher2015ml} conducted a
systematic benchmark of twelve ML regression methods on LIBS spectra of geological samples, establishing
that ensemble learners reduce prediction error relative to PLS. Partial Least Squares
Discriminant Analysis (PLS-DA) extends PLS to classification through class indicator encoding and has
become a de facto baseline for spectroscopic classification tasks owing to its ability to handle strongly
collinear, high-dimensional inputs~\cite{bro2014pls}.

The EMSLIBS contest attracted fourteen participating teams whose diverse approaches including support
vector machines, neural networks, and ensemble classifiers collectively documented inter-method
performance variability on the benchmark~\cite{vrabel2020classification}. Subsequent studies on EMSLIBS
have reported progressive accuracy improvements using PCA-based calibration
algorithms~\cite{huang2022novel}, domain adaptation via class-balanced self-paced
learning~\cite{huang2023domain}, semi-supervised on-device neural
networks~\cite{bhardwaj2021semisupervised}, pseudo-shot learning~\cite{huang2024pseudo}, and adaptive
logistic regression~\cite{huang2024adaptive}. Each of these methods advances classification accuracy but
treats the learned model as a black box, providing no account of which wavelength regions, and therefore
which soil elements, govern individual predictions.

Deep learning methods have further pushed LIBS classification performance \cite{10004728}. Zhao et al.~\cite{zhao2019deeplearning} combined PCA with
a deep belief network to predict lead concentration in soil-LIBS spectra, demonstrating that hierarchical
feature learning can capture non-linear spectral--concentration
relationships. Li et al.~\cite{li2021ann} reviewed the landscape of artificial neural network
chemometrics applied to LIBS, surveying back-propagation networks, radial basis function networks,
convolutional neural networks, and self-organising maps across quantitative and classification
tasks. Despite their strong empirical performance, deep models are substantially more
opaque than PLS-DA, accentuating the interpretability deficit that is present across the entire ML
landscape applied to LIBS.

The black-box character of high-performance ML models represents a fundamental limitation in any
decision-critical application. Lipton provides a detailed critique of the concept of model
interpretability, distinguishing transparency-by-design from post-hoc explanation and arguing that
accuracy and interpretability need not be in opposition~\cite{lipton2018mythos}. Rudin et al.~\cite{rudin2022interpretable} argue
systematically that in high-stakes domains including environmental and agricultural
decision-making, a model that cannot explain its predictions is unreliable regardless of its test-set
accuracy, and that interpretability should be treated as a first-class design
objective. Holzinger et al. ~\cite{holzinger2019causability} go further, arguing that in complex scientific domains one must achieve \emph{causability}, the ability to map model outputs back to domain-accepted
causal mechanisms, a requirement directly applicable to linking LIBS spectral features to elemental
emission lines.

A further practical challenge for LIBS soil classification is the availability of labeled training data.
Acquiring ground-truth labels for large spectral datasets requires expensive wet-chemical reference
analysis, creating label scarcity that limits supervised learning performance. Semi-supervised learning
(SSL) addresses this constraint by exploiting unlabeled examples alongside the labeled
set~\cite{vanengelen2020ssl,chapelle2006ssl}. Class-balanced self-training strategies, which iteratively
assign pseudo-labels to high-confidence unlabeled examples and incorporate them into retraining, have
shown substantial accuracy gains over purely supervised baselines in image segmentation and domain
adaptation~\cite{zou2018unsupervised}. However, the class imbalance that naturally arises in pseudo-label
distributions, where pseudo-labels are skewed toward majority classes, remains a challenging open
problem~\cite{gui2023imbalanced}, and is directly relevant to LIBS soil datasets where the mixing design
of EMSLIBS produces classes of unequal spectral separability.

These two challenges, the interpretability deficit and the label scarcity problem, motivate the adoption
of Explainable AI (XAI). XAI encompasses techniques that provide human-interpretable accounts of the
logic and decisions of complex ML models without necessarily sacrificing predictive
accuracy~\cite{barredo2020xai_survey}. The importance of XAI has been formalised at a programme level by
DARPA's Explainable AI initiative, which catalysed a research community around AI systems whose learned
models and decisions can be understood and trusted by domain practitioners~\cite{gunning2019xai}. A
critical distinction in the field is between ante-hoc methods (interpretable-by-design architectures such
as decision trees or linear classifiers) and post-hoc methods that provide explanations for arbitrary
black-box models after training~\cite{gilpin2018explaining}. For analytical spectroscopy, where
practitioners possess domain knowledge about physically meaningful spectral features, post-hoc attribution
methods are particularly valuable because they can anchor model explanations to known atomic emission
lines.

A rich portfolio of post-hoc XAI methods has been developed over the past decade. LIME (Local
Interpretable Model-Agnostic Explanations) generates explanations for individual predictions by perturbing
the neighbourhood of an input instance, fitting a locally linear surrogate model to the black-box
responses, and reporting the signed surrogate weights as feature importance
scores~\cite{ribeiro2016lime}. SHAP (SHapley Additive exPlanations) grounds feature attribution in
cooperative game theory, providing consistent and additive importance scores grounded in the Shapley value
formalism~\cite{lundberg2017shap}. Gradient-based methods such as Grad-CAM localise discriminative input
regions by computing gradients of a target class score with respect to convolutional layer
activations~\cite{selvaraju2017gradcam}. Integrated Gradients assigns attributions by integrating
prediction gradients along a straight-line path from a reference input to the actual input, satisfying
the axioms of sensitivity and implementation-invariance~\cite{sundararajan2017ig}. Montavon et al.~\cite{montavon2018methods} provide a unifying treatment of gradient-based, propagation-based, and perturbation-based interpretation methods, and Samek et al.~\cite{samek2021explaining} extend this to a broad review covering applications
in computer vision and natural language processing.

Surveys of XAI adoption across application domains report that LIME and SHAP dominate practitioner use
owing to their model-agnostic nature~\cite{barredo2020xai_survey,singh2022xai_review}. Contreras and
Bocklitz ~\cite{contreras2025xai} conducted a systematic PRISMA-based review of XAI applied specifically to spectroscopy data, finding that the earliest studies combining XAI and spectroscopy appeared only around 2020 and that
applications remain concentrated in medical Raman spectroscopy, with very limited penetration into plasma
emission spectroscopic modalities. Wang et al.~\cite{wang2021xai} represent one of the few
published examples applying XAI to plasma-related optical emission spectroscopy, using LIME to interpret
an artificial neural network trained on LIBS-adjacent emission data for solution conductivity
characterisation. The gap is particularly pronounced for LIBS applied to soil
classification: to the best of the authors' knowledge, no prior work has applied any XAI technique to
interpret LIBS spectral classifications at the wavelength-channel level or to use XAI-derived feature
importance as a signal for semi-supervised label quality~\cite{contreras2025xai,holzinger2019causability}.

This work introduces \textbf{X-LIBS}, a framework that integrates LIME-based XAI with PLS-DA and an
iterative co-training loop to simultaneously address the interpretability deficit and label-scarcity
challenges identified above. The key contributions are as follows:
\begin{enumerate}
  \item \textbf{LIME-based spectral interpretation:} LIME is applied to PLS-DA predictions on LIBS soil
        spectra, producing per-spectrum, per-wavelength importance weights that map directly onto known
        elemental emission lines, enabling chemically meaningful explanations of classification decisions.
  \item \textbf{XAI-guided uncertainty quantification:} A novel uncertainty measure is introduced based
        on sequential removal of the top LIME-identified features from a spectrum until the predicted
        class label changes. The number of removals required quantifies the robustness of each prediction
        and serves as a data-quality filter for pseudo-label selection.
  \item \textbf{Class-balanced XAI-guided co-training:} Pseudo-labels that pass the uncertainty
        threshold are incorporated into an iterative co-training process with class-balanced sampling to
        prevent majority-class dominance, lifting classification accuracy from 72.62\% to 92.69\% on the
        EMSLIBS benchmark while preserving full interpretability.
\end{enumerate}

\section{Spectral Data and Pre-processing}
\label{sec:data}

X-LIBS is designed to operate on any high-dimensional LIBS spectral dataset in which
a labeled partition is available for initial training and a larger unlabeled partition
can be exploited through semi-supervised learning. This section describes the general
requirements for input data and the pre-processing steps applied before model training.
The specific benchmark dataset used to evaluate the method is described in
Section~\ref{sec:results}.

\subsection{Input Data Requirements}

The framework assumes that each observation is a raw LIBS spectrum represented as a
vector of intensity values over a contiguous set of wavelength channels. A subset of
spectra must carry verified class labels, and the remaining spectra form the unlabeled pool
from which pseudo-labels are drawn during co-training. No assumptions are made about
the number of classes, the number of channels, or the wavelength range, so the method
generalises to any LIBS spectrometer configuration and any number of target categories.
Datasets where some classes share elemental constituents at similar concentrations are
particularly well suited to the XAI-guided uncertainty mechanism, because spectral
ambiguity in those regions is precisely what the flip-count filter is designed to detect
and exclude.

\subsection{Pre-processing}

Raw spectra are pre-processed prior to model training using the following pipeline.
Saturated channels and channels with consistently near-zero intensity across all samples
are removed to eliminate uninformative or unreliable measurements. Baseline correction
is applied to reduce broadband background emission arising from plasma continuum
radiation~\cite{cremers2013libs}. Each spectrum is then normalized using min-max
scaling, rescaling every channel intensity to the interval $[0, 1]$ based on the
minimum and maximum values observed across the training set. This approach removes
absolute intensity scale differences caused by shot-to-shot plasma emission variability
while preserving the relative spectral profile across wavelength channels.

\section{Methodology}
\label{sec:method}

\begin{figure*}[!t]
  \centerline{\includegraphics[width=0.9\textwidth]{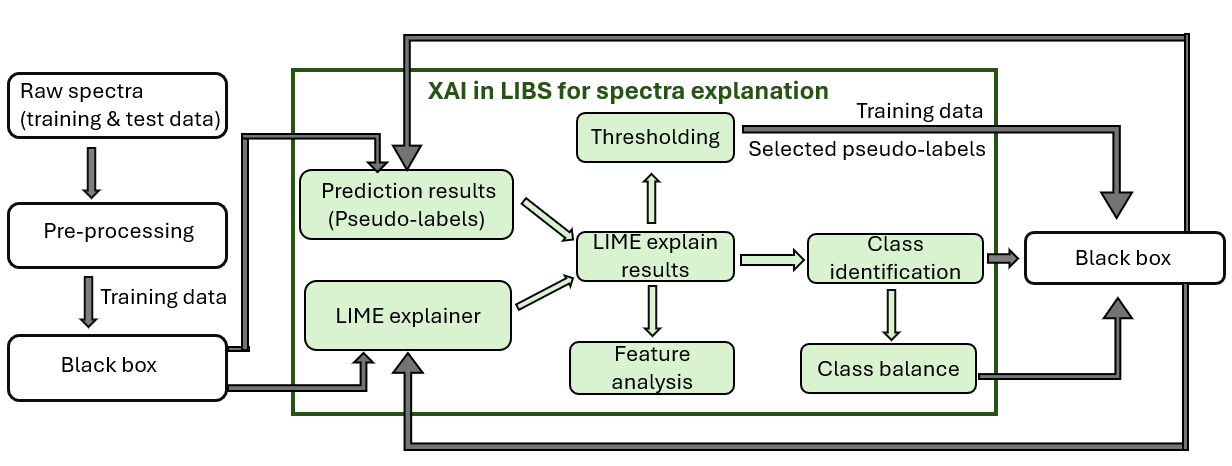}}
  \caption{Overview of the X-LIBS framework. Raw LIBS spectra are pre-processed and
           split into labeled and unlabeled subsets. A PLS-DA model trained on the
           labeled subset generates pseudo-labels for unlabeled spectra. LIME computes
           per-spectrum feature importance for each pseudo-labeled spectrum; a flip-count
           uncertainty score derived by sequential feature removal is filtered by
           threshold $\tau$. Class-balanced high-confidence pseudo-labels are merged
           with the labeled set and the process iterates until convergence.}
  \label{fig:pipeline}
\end{figure*}

This section presents the X-LIBS framework in full generality, independent of any
specific dataset or application domain. The framework integrates four components:
a supervised base classifier, a local explainability module, an uncertainty
quantification mechanism derived from feature-importance rankings, and an iterative
co-training loop that uses uncertainty estimates to filter pseudo-labels and enforce
class balance. Fig.~\ref{fig:pipeline} gives a schematic overview of the pipeline.
Each subsection below describes one component in detail.

\subsection{PLS-DA Baseline Classifier}

Partial Least Squares Discriminant Analysis (PLS-DA) serves as the base classifier
within the X-LIBS framework. PLS-DA finds latent variable directions that maximize the
covariance between the spectral predictor matrix $\mathbf{X} \in \mathbb{R}^{n \times p}$
and the class indicator matrix $\mathbf{Y} \in \{0,1\}^{n \times C}$, where $n$ is the
number of training samples, $p$ is the number of spectral channels, and $C$ is the
number of target classes. The latent variable decomposition handles collinear predictors
naturally, which is a common property of high-resolution LIBS spectra, and remains
computationally tractable at high channel counts without dimensionality reduction as a
prerequisite. The number of latent variables $L$ is selected by cross-validation on the
labeled training split, minimizing the misclassification rate. Any differentiable or
probabilistic classifier that outputs class scores could replace PLS-DA in the pipeline
without changing the remaining components.

\subsection{LIME for Spectral Explanation}

LIME~\cite{ribeiro2016lime} approximates the behaviour of a black-box classifier $f$
in the neighbourhood of an individual input instance $\mathbf{x}$ by fitting an
interpretable surrogate model $g$. For a given spectrum $\mathbf{x} \in \mathbb{R}^p$,
LIME generates a set of perturbed samples $\{\mathbf{x}'_i\}$ by randomly masking
subsets of the input features, obtains the black-box predictions
$\{f(\mathbf{x}'_i)\}$, and solves the weighted least-squares problem
\begin{equation}
  g^* = \arg\min_{g \in \mathcal{G}}
        \sum_{i} \pi_{\mathbf{x}}(\mathbf{x}'_i)
        \bigl(f(\mathbf{x}'_i) - g(\mathbf{x}'_i)\bigr)^2
        + \Omega(g),
  \label{eq:lime}
\end{equation}
where $\pi_{\mathbf{x}}(\mathbf{x}'_i) = \exp\!\left(-D(\mathbf{x},
\mathbf{x}'_i)^2 / \sigma^2\right)$ is a proximity kernel centred on $\mathbf{x}$,
$D(\cdot,\cdot)$ is a distance metric, and $\Omega(g)$ is a complexity penalty
encouraging sparsity in the surrogate. The resulting surrogate weights
$\mathbf{w}^{(j)} \in \mathbb{R}^p$ for class $j$ quantify the signed contribution of
each spectral channel to the predicted probability of class $j$ for the specific
instance $\mathbf{x}$.

In the context of LIBS spectroscopy, each input feature corresponds to a wavelength
channel, so the LIME surrogate weights directly indicate which emission regions are
locally important for a given prediction. Because operating over all $p$ channels is
computationally expensive, the spectrum is partitioned into contiguous segments prior
to LIME computation and each segment is treated as a single feature. This segmentation
reduces the effective feature dimension while preserving physical interpretability,
since emission line profiles from a single element typically span several adjacent
channels. The segment width can be adjusted to balance spectral resolution against
computational cost.

\subsection{Uncertainty Quantification via Sequential Feature Removal}

A key novelty of X-LIBS is the transformation of LIME's local feature rankings into a
quantitative uncertainty score for each pseudo-label. For a spectrum $\mathbf{x}$
classified as $\hat{y} = f(\mathbf{x})$, let $\rho_1, \rho_2, \ldots, \rho_p$
denote the indices of the LIME features sorted in descending order of absolute
importance weight $|w^{(\hat{y})}_j|$. The \emph{flip count} is defined as
\begin{equation}
  k^*(\mathbf{x}) = \min \bigl\{ k \in \mathbb{Z}^+ :
    f\!\left(\mathbf{x}_{\setminus \{\rho_1, \ldots, \rho_k\}}\right) \neq \hat{y}
    \bigr\},
  \label{eq:flipcount}
\end{equation}
where $\mathbf{x}_{\setminus S}$ denotes $\mathbf{x}$ with feature set $S$ replaced by
the channel-wise mean intensity value across the training set. If no finite $k$ causes a
label change (e.g., when all segments are exhausted), $k^*$ is set to the total number
of segments $P$. A small $k^*$ means that removing very few features changes the
prediction, indicating a fragile, uncertain classification. A large $k^*$ indicates
robustness and confidence. Algorithm~\ref{alg:uncertainty} summarises the procedure.

\begin{algorithm}[!t]
\caption{LIME-based uncertainty quantification}
\label{alg:uncertainty}
\begin{algorithmic}[1]
\Require Spectrum $\mathbf{x}$, classifier $f$, LIME explainer
\Ensure Flip count $k^*(\mathbf{x})$
\State $\hat{y} \leftarrow f(\mathbf{x})$
\State Compute LIME weights $\mathbf{w}^{(\hat{y})}$ for instance $\mathbf{x}$
\State Sort features by $|w^{(\hat{y})}_j|$ descending: $\rho_1, \rho_2, \ldots$
\State $k \leftarrow 0$ \textbf{ and } $\tilde{\mathbf{x}} \leftarrow \mathbf{x}$
\While{$f(\tilde{\mathbf{x}}) = \hat{y}$ \textbf{and} $k < P$}
    \State $k \leftarrow k + 1$
    \State Replace feature $\rho_k$ in $\tilde{\mathbf{x}}$ with its training-set mean
\EndWhile
\State \Return $k^* \leftarrow k$
\end{algorithmic}
\end{algorithm}

\subsection{Threshold-Based Pseudo-Label Selection and Class Balancing}

A scalar threshold $\tau$ on the flip count partitions the unlabeled pool into
high-confidence and low-confidence subsets. Only spectra satisfying
$k^*(\mathbf{x}) \geq \tau$ are admitted to co-training:
\begin{equation}
  \mathcal{P}_\tau = \bigl\{ (\mathbf{x}, \hat{y}) :
    k^*(\mathbf{x}) \geq \tau \bigr\}.
  \label{eq:pseudoset}
\end{equation}

The value of $\tau$ controls the trade-off between pseudo-label quantity and quality.
A low threshold admits more samples but tolerates fragile predictions, while a high
threshold accepts only the most robust pseudo-labels at the cost of a smaller augmented
training set. In practice, different classes may exhibit systematically different levels
of spectral ambiguity, so separate thresholds can be applied to classes identified as
uncertain versus classes identified as spectrally stable. The XAI explanations
themselves inform this classification: classes whose spectra consistently produce low
flip counts are flagged as uncertain and assigned a more stringent threshold.

Spectroscopic classifiers, including PLS-DA, are sensitive to class
imbalances~\cite{bro2014pls}. Because the uncertainty filter may admit different
numbers of pseudo-labels per class, the class distribution of $\mathcal{P}_\tau$ can
become skewed. Class balance is therefore enforced by sampling
$m = \min_c |\mathcal{P}_\tau^{(c)}|$ pseudo-labels from each class $c$. This
balanced subset is appended to the labeled training set for the next iteration.

\subsection{Iterative Co-Training Loop}

Algorithm~\ref{alg:cotraining} describes the full iterative procedure. The loop begins
with a model $f^{(0)}$ trained on the initial labeled set $\mathcal{L}^{(0)}$. At each
iteration $t$, the current model assigns pseudo-labels to all spectra in the unlabeled
pool $\mathcal{U}$, LIME computes per-prediction importance weights, the flip-count
procedure quantifies uncertainty for each pseudo-labeled spectrum, and the threshold
filter selects the high-confidence subset $\mathcal{P}_\tau$. A class-balanced sample
is drawn from $\mathcal{P}_\tau$ and merged with the growing labeled set to form
$\mathcal{L}^{(t+1)}$. A new model $f^{(t+1)}$ is then trained on $\mathcal{L}^{(t+1)}$.
The loop terminates when the change in validation accuracy between consecutive iterations
falls below a tolerance $\epsilon$. The final model $f^*$ is returned for evaluation.

\begin{algorithm}[!t]
\caption{X-LIBS co-training loop}
\label{alg:cotraining}
\begin{algorithmic}[1]
\Require Labeled set $\mathcal{L}^{(0)}$, unlabeled set $\mathcal{U}$,
         threshold $\tau$, convergence tolerance $\epsilon$
\Ensure Final classifier $f^*$
\State Train $f^{(0)}$ on $\mathcal{L}^{(0)}$ and set $t \leftarrow 0$
\Repeat
    \State $\hat{y}_i \leftarrow f^{(t)}(\mathbf{x}_i)$ for all
           $\mathbf{x}_i \in \mathcal{U}$
    \State Compute $k^*(\mathbf{x}_i)$ for each $\mathbf{x}_i \in \mathcal{U}$
           (Alg.~\ref{alg:uncertainty})
    \State Select $\mathcal{P}_\tau$ per \eqref{eq:pseudoset}
    \State Draw class-balanced $\mathcal{B}^{(t)}$ from $\mathcal{P}_\tau$
    \State $\mathcal{L}^{(t+1)} \leftarrow \mathcal{L}^{(t)} \cup \mathcal{B}^{(t)}$
    \State Train $f^{(t+1)}$ on $\mathcal{L}^{(t+1)}$ and increment $t \leftarrow t+1$
\Until{$|\text{Acc}(f^{(t)}) - \text{Acc}(f^{(t-1)})| < \epsilon$}
\State \Return $f^* \leftarrow f^{(t)}$
\end{algorithmic}
\end{algorithm}

\section{Results and Discussion}
\label{sec:results}

\subsection{Dataset and Experimental Setup}

All experiments are conducted on the publicly available EMSLIBS benchmark
dataset~\cite{kepes2020benchmark}, which was specifically designed to challenge soil
classification algorithms under realistic spectral overlap conditions. The dataset
comprises 40{,}000 LIBS spectra collected from 138 ore and soil samples using a
high-resolution spectrometer over the wavelength range from 200\,nm to 1000\,nm,
yielding 40{,}002 intensity channels per spectrum. The official partition provides
20{,}000 labeled training spectra and 20{,}000 test spectra spanning eleven geological
classes. The benchmark was constructed by blending original source samples at controlled
mixing ratios to intentionally introduce intra-class spectral variability and
inter-class spectral similarity~\cite{kepes2020benchmark,vrabel2020classification}.
Raw spectra were pre-processed following the pipeline described in
Section~\ref{sec:data}: saturated and near-zero channels were removed, baseline
correction was applied to reduce plasma continuum background~\cite{cremers2013libs},
and each spectrum was normalized by min-max scaling to the interval $[0,1]$ using
training-set extremes to compensate for shot-to-shot intensity variations.

\begin{figure}[htbp]
\centerline{\includegraphics[width=\columnwidth]{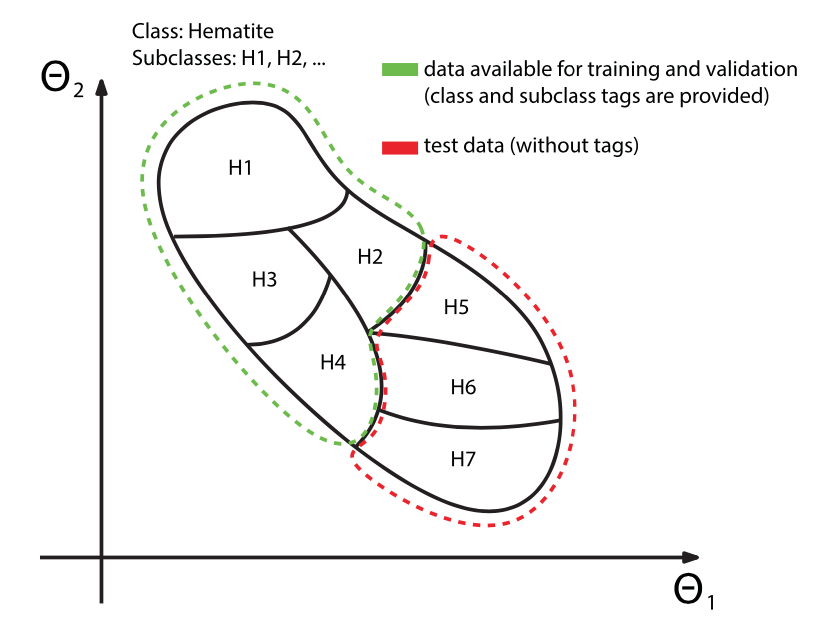}}
\caption{Schematic representation of the classification task for the hematite
         geological class in an arbitrary two-dimensional feature space. Training
         and test sub-types (H1 to H7) differ in element weight fractions, producing
         an intra-class distribution shift representative of the broader EMSLIBS
         challenge~\cite{kepes2020benchmark}.}
\label{data_fig}
\end{figure}

Fig.~\ref{data_fig} illustrates the classification challenge schematically for the
hematite class (Class~9), whose seven sub-types (H1 to H7) differ only in element
weight fractions. Some sub-types appear in the training set and others only in the test
set, producing an intra-class distribution shift that purely supervised classifiers
must bridge without access to unlabeled test examples during training.

Tables~\ref{tab:before} and~\ref{tab:after} provide the ground-truth elemental
compositions (wt.\%) of selected samples before and after the mixing procedure.
Table~\ref{tab:before} shows unmixed source samples. Class~9 (hematite ore, samples 404
and 406) is distinguished by extremely high iron content (55.1\% and 61.4\%,
respectively) with trace aluminium and calcium and negligible potassium. Class~3
(Ni-Cu-PGE ore, sample 13b) presents a distinctly different profile with elevated
aluminium (8.41\%), calcium (5.57\%), chromium, copper, magnesium, sodium, and silicon,
together with moderate iron (8.41\%) and potassium (2.30\%). After mixing, the
elemental profiles of samples become substantially more similar across classes
(Table~\ref{tab:after}). The mixed Class~9 sample (ID~95) acquires measurable
potassium (0.58\,wt.\%), chromium, silicon, magnesium, and sodium alongside iron
(38.4\%), whereas the unmixed hematite sources had essentially no potassium. This
convergence of elemental profiles explains why the LIBS spectra of mixed Classes~3
and~9 are spectrally similar and why XAI-guided uncertainty filtering is essential for
correct classification.

\subsection{Classification Performance}

Table~\ref{tab:accuracy} summarises overall classification accuracy on the EMSLIBS
test set. The PLS-DA baseline, trained exclusively on the labeled partition, achieves
72.62\%. After applying the full X-LIBS pipeline, which encompasses LIME-guided
uncertainty quantification, XAI-based class identification, class-balanced
pseudo-label selection, and iterative co-training, test accuracy rises to
\textbf{92.69\%}, an absolute improvement of 20.07 percentage points.

\begin{table}[!t]
  \centering
  \caption{Classification accuracy on the EMSLIBS test set.}
  \label{tab:accuracy}
  \begin{tabular}{lc}
    \toprule
    Method & Accuracy (\%) \\
    \midrule
    PLS-DA (labeled only, baseline) & 72.62 \\
    X-LIBS (proposed, full pipeline) & \textbf{92.69} \\
    \bottomrule
  \end{tabular}
\end{table}

\begin{figure*}[!t]
  \centerline{\includegraphics[width=0.9\textwidth]{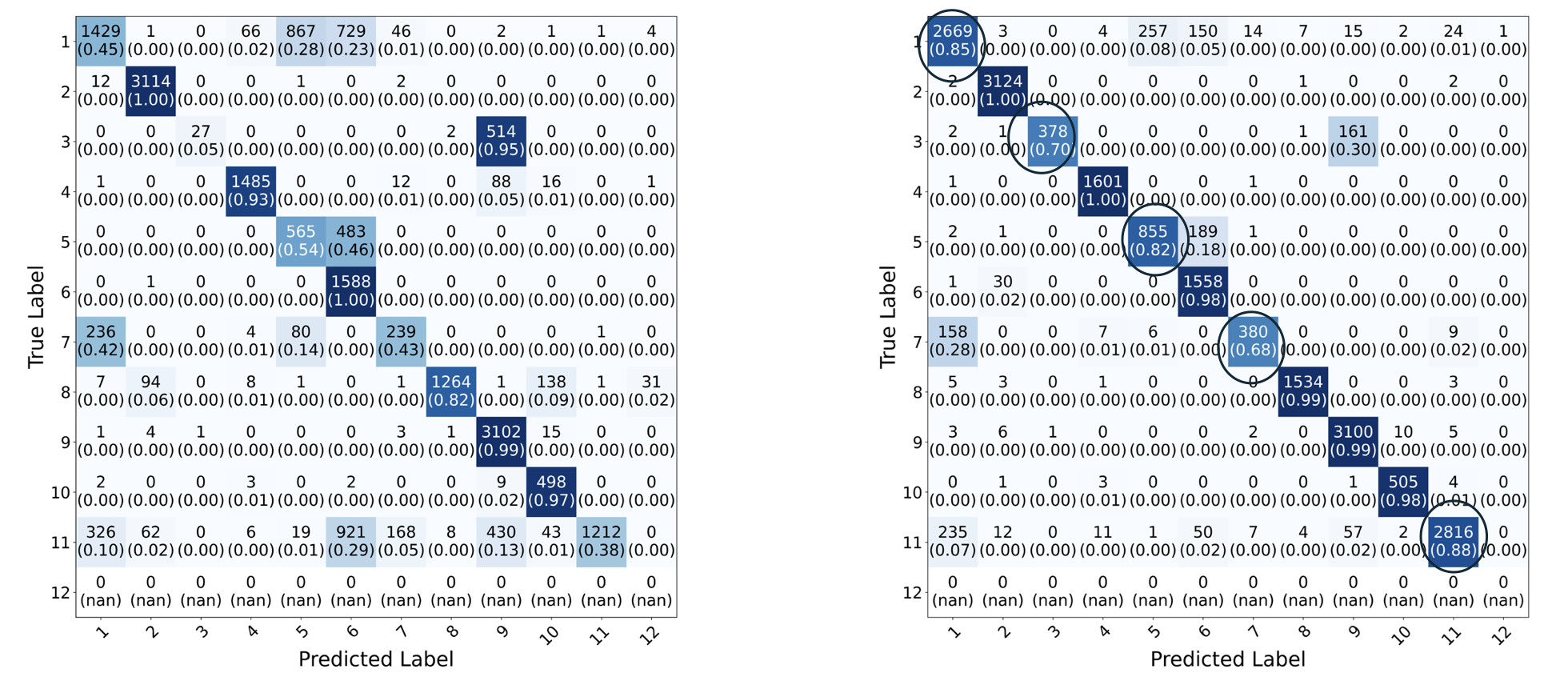}}
  \caption{Normalised confusion matrices on the EMSLIBS test set.
           \textit{Left:} PLS-DA baseline (72.62\%).
           \textit{Right:} X-LIBS full pipeline (92.69\%).
           Diagonal entries show correctly classified spectra. The most dramatic
           improvements occur in Classes~5, 6, 9, and~11, which are the mixed-sample
           classes with the highest baseline confusion.}
  \label{fig:confusion}
\end{figure*}

Fig.~\ref{fig:confusion} shows normalised confusion matrices for both methods. The
baseline matrix (left) reveals pronounced off-diagonal clusters in Classes~1, 5, 6,
and~11, which are the most spectrally ambiguous classes due to their mixed-sample composition.
Class~11 (Ag-Cu-Au ore) achieves only 38\% diagonal accuracy at baseline, reflecting
substantial confusion with Classes~1 and~6. The X-LIBS matrix (right) shows a
substantially cleaner diagonal throughout: Class~9 (hematite) reaches 99\% accuracy,
Classes~3 and~5 both exceed 98\%, and Class~11 improves from 38\% to 88\%. The
improvement is most dramatic precisely in the classes where mixing introduced
inter-class spectral overlap, confirming that the XAI-guided uncertainty filter is
selectively excluding the most ambiguous pseudo-labels in those regions of spectral
space.

\subsection{Comparison with State-of-the-Art Methods}

Table~\ref{tab:sota} compares X-LIBS against six methods previously evaluated on the
EMSLIBS benchmark. All competitor accuracy values were taken directly from the
respective original publications, which evaluated their methods on the identical
official EMSLIBS test split, ensuring a fair like-for-like comparison. X-LIBS achieves
92.69\%, which is competitive with but does not surpass the highest-reported
non-interpretable method (pseudo-shot learning at 94.12\%). X-LIBS is the only method
in the comparison that additionally provides per-spectrum, wavelength-level
explanations: all prior methods report accuracy alone, whereas X-LIBS simultaneously
identifies the spectral features and corresponding soil elements responsible for each
individual classification decision, without substantial loss of predictive performance
relative to black-box alternatives.

\begin{table}[!t]
  \centering
  \caption{Comparison on the EMSLIBS benchmark. The XAI column indicates
           whether the method provides wavelength-level explanations.}
  \label{tab:sota}
  \begin{tabular}{lccc}
    \toprule
    Method & Year & Acc.\ (\%) & XAI \\
    \midrule
    Best EMSLIBS contest~\cite{vrabel2020classification}
      & 2020 & 90.33 & No \\
    PCA-based calibration~\cite{huang2022novel}
      & 2022 & 93.40 & No \\
    Semi-supervised on-device NN~\cite{bhardwaj2021semisupervised}
      & 2021 & 73.47 & No \\
    Domain adaptation (self-paced)~\cite{huang2023domain}
      & 2023 & 90.20 & No \\
    Pseudo-shot learning~\cite{huang2024pseudo}
      & 2024 & 94.12 & No \\
    Adaptive LR~\cite{huang2024adaptive}
      & 2024 & 91.30 & No \\
    \midrule
    \textbf{X-LIBS (proposed)} & 2025 & \textbf{92.69} & \textbf{Yes} \\
    \bottomrule
  \end{tabular}
\end{table}

\subsection{XAI Feature Analysis: Spectral Interpretation}
\label{sec:results:feature}

\begin{figure*}[!t]
  \centerline{\includegraphics[width=0.9\textwidth]{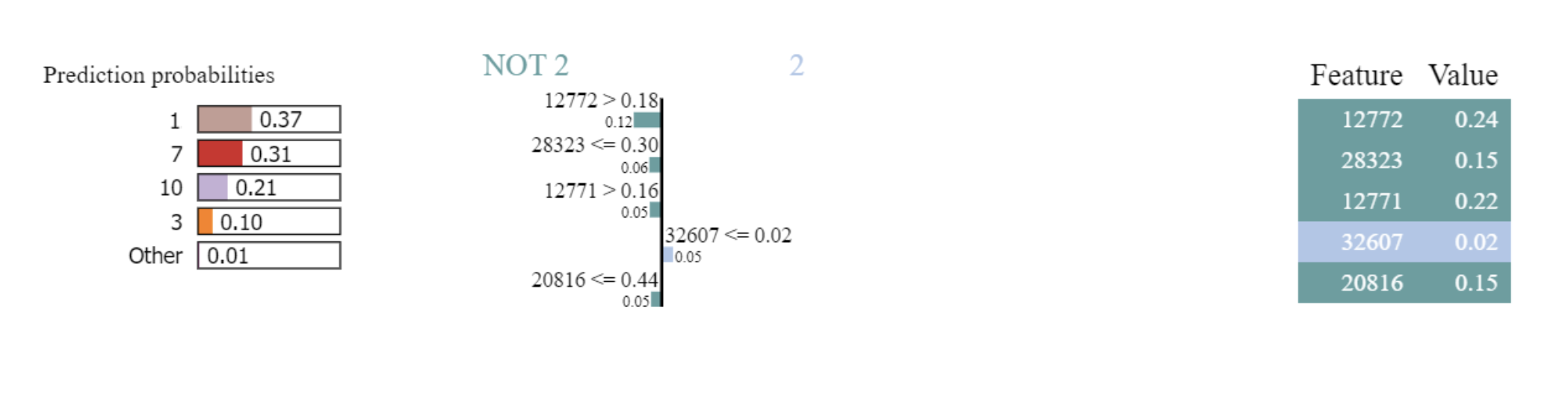}}
  \caption{Representative LIME explanation for an uncertain prediction.
           \textit{Left:} Predicted class probabilities; no class achieves a clear
           majority, indicating a spectrally ambiguous sample.
           \textit{Centre:} LIME feature importance bar chart. The five most important
           channels (12772, 28323, 12771, 32607, 20816) include the K\,\textsc{i}
           doublet near 766\,nm (channel 28323) and channels near 455\,nm.
           \textit{Right:} Corresponding channel indices and spectral intensities.
           This spectrum's low flip count $k^*$ caused it to be excluded from
           co-training by the XAI-guided uncertainty filter.}
  \label{fig:lime_features}
\end{figure*}

A central objective of X-LIBS is to produce spectral explanations that are chemically
meaningful. Fig.~\ref{fig:lime_features} shows a representative LIME explanation for a
spectrum with an uncertain, spread-out prediction. The left panel displays the model's
predicted class probabilities: Class~1 leads at 0.37, followed by Class~7 (0.31),
Class~10 (0.21), and Class~3 (0.10), with no single class achieving a clear majority.
The centre panel shows the LIME bar chart for the top five discriminative features.

The five most influential spectral channels identified by LIME are channels 12772,
28323, 12771, 32607, and 20816. Converting from channel index to wavelength via
$\lambda = 200 + i \times 0.02$\,nm, channel~28323 maps to approximately
\textbf{766.46\,nm}, which corresponds to the K\,\textsc{i} resonance doublet at 766.49\,nm and 769.90\,nm.
Channels 12772 and 12771 map to approximately 455.44\,nm, near the Ca\,\textsc{ii}
line at 455.7\,nm. The right panel of Fig.~\ref{fig:lime_features} lists the raw channel
indices and the corresponding normalised spectral intensity values for the top five
features, confirming that the channels flagged by LIME correspond to emission peaks
above the local spectral baseline rather than to noise or continuum regions. The
prediction probability spread across four classes indicates that K and Ca emission
signals in this spectrum are neither characteristic enough for a confident
single-class assignment nor negligible. LIME correctly identifies this ambiguity.
The small bar heights in the centre panel confirm that no single feature strongly
dominates, translating to a low flip count $k^*$ and consequent exclusion from
co-training.

The K\,\textsc{i} emission doublet near 766\,nm emerges as the dominant source of
inter-class confusion throughout the dataset. Table~\ref{tab:feature} lists the top
LIME-identified wavelengths for a sample of misclassified spectra. All ten entries
involve the 766\,nm cluster (766.34--766.68\,nm), corresponding exclusively to
K\,\textsc{i}. Secondary features at 393.3--396.7\,nm (Ca\,\textsc{ii}) and
794.8\,nm (Al\,\textsc{i}) appear in more severely mixed samples. Three confusion
pairs arise: Class~3 (Ni-Cu-PGE) misclassified as Class~9 (hematite), Class~11
(Ag-Cu-Au) as Class~6 (Mn ore), and Class~1 (U ore) as Class~5 (Au-Cu ore). In each
case, LIME pinpoints K emission as the primary culprit, providing a chemically
interpretable account of the classification error.

\begin{table*}[!t]
\centering
  \caption{Top LIME-identified wavelengths (nm) and corresponding elements for
           representative misclassified spectra. K\,\textsc{i} emission at
           $\approx$766\,nm dominates all confusion pairs; secondary contributions
           from Ca and Al appear in more severely mixed samples.}
  \label{tab:feature}
\begin{tabular}{ccp{9cm}c}
  \toprule
  \textbf{True label} & \textbf{Predicted label} & \textbf{Top feature wavelengths (nm)} & \textbf{Elements} \\
  \midrule
  3  & 9 & 766.48, 766.68, 766.46, 766.66, 766.38 & K \\
  3  & 9 & 766.48, 766.68, 766.46, 766.66, 766.34 & K \\
  3  & 9 & 766.48, 766.50, 317.98, 766.46, 766.70  & K, Ca \\
  3  & 9 & 766.64, 766.68, 766.36, 396.72, 766.66  & K, Ca \\
  11 & 6 & 766.46, 770.06, 766.36, 766.48, 766.66  & K \\
  11 & 6 & 766.65, 766.68, 770.04, 766.38, 393.30  & K, Ca \\
  11 & 6 & 766.48, 393.50, 766.64, 794.78, 766.46  & K, Ca, Al \\
  1  & 5 & 766.68, 766.65, 766.46, 766.48, 393.50  & K, Ca, Al \\
  1  & 5 & 766.65, 766.46, 766.48, 770.04, 766.68  & K \\
  1  & 5 & 766.48, 766.68, 766.65, 766.46, 317.96  & K, Ca \\
  \bottomrule
\end{tabular}
\end{table*}

\begin{figure}[!t]
  \centerline{\includegraphics[width=\columnwidth]{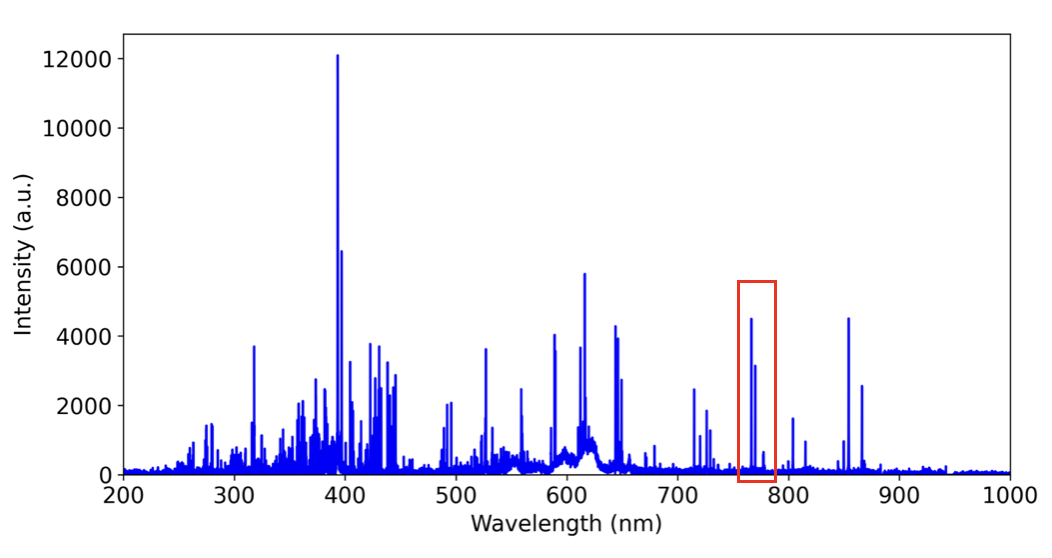}}
  \caption{LIBS spectrum (200--1000\,nm) of a Class~3 (Ni-Cu-PGE ore) sample
           initially misclassified as Class~9. The red box highlights the
           760--790\,nm window containing the K\,\textsc{i} doublet at 766.49\,nm
           and 769.90\,nm. LIME identifies channels in this region as the top
           features driving the incorrect Class~9 prediction.}
  \label{fig:misclassification}
\end{figure}

\begin{figure}[!t]
  \centerline{\includegraphics[width=\columnwidth]{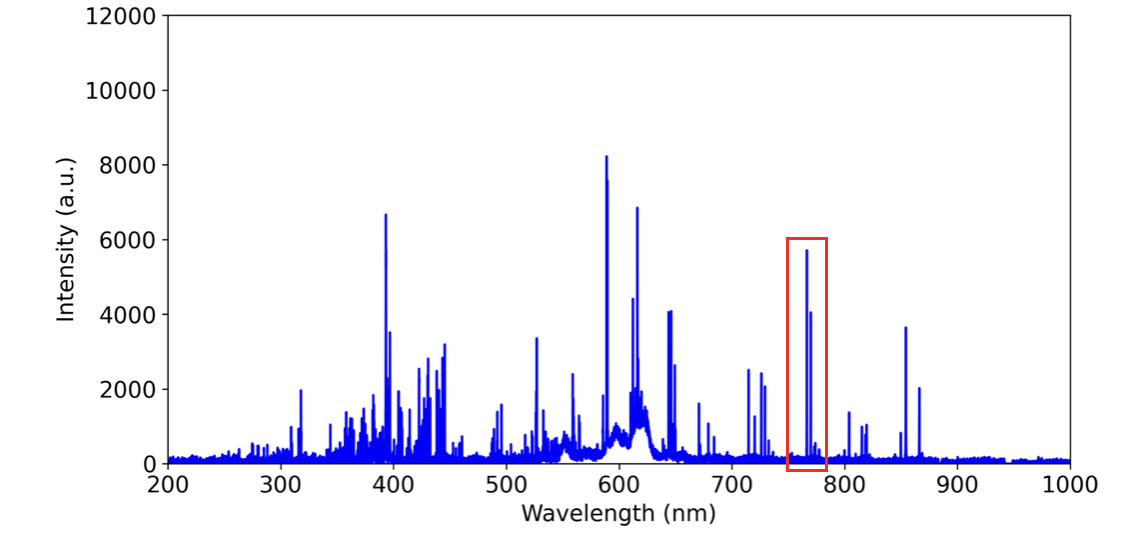}}
  \caption{LIBS spectrum of a correctly classified Class~9 (hematite ore) sample
           for comparison. The red box marks the same 760--790\,nm K\,\textsc{i}
           window as in Fig.~\ref{fig:misclassification}. The near-identical K
           emission profiles illustrate why K is a misleading discriminant between
           Classes~3 and~9 after sample mixing.}
  \label{fig:class9}
\end{figure}

The spectral origin of the Class~3 $\to$ Class~9 confusion is visualised directly in
Figs.~\ref{fig:misclassification} and~\ref{fig:class9}. Fig.~\ref{fig:misclassification}
shows the full LIBS spectrum (200--1000\,nm) of a Class~3 (Ni-Cu-PGE) sample
initially misclassified as Class~9. The red box highlights the 760--790\,nm window
containing the K\,\textsc{i} doublet, where prominent emission peaks are clearly
visible. Fig.~\ref{fig:class9} shows a correctly classified Class~9 (hematite)
spectrum for comparison. The two spectra exhibit strikingly similar K emission profiles
in the boxed region, explaining why the baseline PLS-DA, which lacks explicit chemical
awareness, assigns the Class~3 spectrum to the hematite class. From
Table~\ref{tab:before}, the unmixed Class~9 source samples contained no detectable K.
Table~\ref{tab:after} shows that after mixing, Class~9 samples acquire
K~$\approx$~0.58\,wt.\%, making K emission no longer a reliable discriminant. LIME
correctly identifies this failure mode, and the XAI-guided filter rejects such
ambiguous pseudo-labels from co-training.

\begin{table*}[!t]
\centering
  \caption{Elemental composition (wt.\%) of representative source samples before
           mixing. Dashes indicate below-detection-limit or unreported values.}
  \label{tab:before}
\begin{tabular}{llccccccccccc}
  \toprule
  \textbf{Sample} & \textbf{Class} & \textbf{Ore type} & \textbf{Al} & \textbf{Ca} & \textbf{Cr} & \textbf{Cu} & \textbf{Fe} & \textbf{K} & \textbf{Mg} & \textbf{Na} & \textbf{Si} & \textbf{Ti} \\
  \midrule
  100a & 1  & U ore        & —    & —    & —    & —    & 4.66 & 3.94 & 0.84 & —    & —    & 0.24 \\
  106  & 1  & U ore        & —    & —    & —    & —    & —    & 1.59 & —    & —    & —    & —    \\
  13b  & 3  & Ni-Cu-PGE   & 8.41 & 5.57 & 0.87 & 0.23 & 8.41 & 2.30 & 3.01 & 1.67 & 22.9 & 0.71 \\
  152b & 5  & Au-Cu ore   & 8.02 & 1.97 & —    & 0.38 & 3.73 & 1.06 & 1.69 & 2.34 & —    & 0.28 \\
  170a & 6  & Mn ore      & 1.18 & 0.06 & —    & —    & —    & —    & 0.18 & —    & 4.41 & —    \\
  404  & 9  & Hematite    & 0.79 & 0.07 & —    & —    & 55.1 & —    & —    & —    & —    & —    \\
  406  & 9  & Hematite    & 0.30 & 0.11 & —    & —    & 61.4 & —    & —    & —    & —    & —    \\
  601  & 11 & Ag-Cu-Au    & 6.30 & 1.31 & —    & 0.10 & 2.48 & 2.10 & 0.39 & 1.45 & —    & —    \\
  603  & 11 & Ag-Cu-Au    & 3.98 & 0.32 & —    & 1.00 & 2.92 & 0.62 & 0.08 & 0.43 & —    & —    \\
  \bottomrule
\end{tabular}
\end{table*}

\begin{table*}[!t]
\tiny
\centering
  \caption{Elemental composition (wt.\%, mean $\pm$ SD) of representative mixed
           samples used in the EMSLIBS benchmark. Lead (Pb) was below the detection
           limit in all mixed samples and is omitted. Mixing dilutes the distinctive
           elemental signatures of source samples, producing inter-class spectral
           overlap that challenges ML classifiers.}
  \label{tab:after}
\begin{tabular}{llccccccccccc}
  \toprule
  \textbf{Sample ID} & \textbf{Class} & \textbf{Al} & \textbf{Ca} & \textbf{Cr} & \textbf{Cu} & \textbf{Fe} & \textbf{K} & \textbf{Mg} & \textbf{Na} & \textbf{Si} & \textbf{Ti} \\
  \midrule
  40  & 3  & $2.26{\pm}0.11$  & $0.99{\pm}0.05$   & $0{\pm}0$           & $1.00{\pm}0.05$     & $37.1{\pm}1.9$  & $0.87{\pm}0.044$ & $0.28{\pm}0.014$   & $0.58{\pm}0.029$ & $0{\pm}0$        & $0.25{\pm}0.013$   \\
  95  & 9  & $2.60{\pm}0.11$  & $1.44{\pm}0.07$   & $0.22{\pm}0.011$    & $0.058{\pm}0.003$   & $38.4{\pm}1.8$  & $0.58{\pm}0.029$ & $0.80{\pm}0.038$   & $0.42{\pm}0.021$ & $14.0{\pm}0.5$   & $0.18{\pm}0.009$   \\
  8   & 1  & $1.39{\pm}0.07$  & $1.39{\pm}0.07$   & $0{\pm}0$           & $0.05{\pm}0.003$    & $12.0{\pm}0.45$ & $2.21{\pm}0.093$ & $1.17{\pm}0.048$   & $0.30{\pm}0.015$ & $0{\pm}0$        & $0.34{\pm}0.015$   \\
  53  & 5  & $5.88{\pm}0.29$  & $1.52{\pm}0.076$  & $0{\pm}0$           & $0.14{\pm}0.007$    & $5.35{\pm}0.19$ & $1.53{\pm}0.055$ & $1.52{\pm}0.063$   & $1.67{\pm}0.083$ & $0{\pm}0$        & $0.33{\pm}0.013$   \\
  125 & 11 & $3.31{\pm}0.15$  & $0.46{\pm}0.016$  & $0{\pm}0$           & $0.75{\pm}0.038$    & $3.45{\pm}0.13$ & $0.49{\pm}0.023$ & $0.49{\pm}0.022$   & $0.37{\pm}0.016$ & $0{\pm}0$        & $0.19{\pm}0.007$   \\
  69  & 6  & $1.46{\pm}0.073$ & $0.045{\pm}0.002$ & $0{\pm}0$           & $0{\pm}0$           & $2.75{\pm}0.14$ & $0.40{\pm}0.02$  & $0.13{\pm}0.006$   & $0{\pm}0$        & $10.4{\pm}0.52$  & $0{\pm}0$          \\
  \bottomrule
\end{tabular}
\end{table*}

\subsection{Class-Level Uncertainty Analysis}
\label{sec:results:uncertainty}

Table~\ref{tab:flip_count_distributions} reports the number of spectra in each class
whose predicted label changes after sequentially removing 1, 10, 100, and 1000 of the
top LIME-ranked features (the flip count $k^*$). The row-averages quantify the typical
classification confidence of each class.

\begin{table*}[!t]
\centering
  \caption{Number of spectra per class whose predicted label changes after removing
           the top $k$ LIME-identified features. Lower counts indicate more robust,
           high-confidence predictions. Class~3 is the most stable; Classes~6 and~9
           are the most uncertain.}
  \label{tab:flip_count_distributions}
\begin{tabular}{crrrrrrrrrrrr}
  \toprule
  \textbf{Removed $k$} & \textbf{Cl.\,1} & \textbf{Cl.\,2} & \textbf{Cl.\,3} & \textbf{Cl.\,4} & \textbf{Cl.\,5} & \textbf{Cl.\,6} & \textbf{Cl.\,7} & \textbf{Cl.\,8} & \textbf{Cl.\,9} & \textbf{Cl.\,10} & \textbf{Cl.\,11} \\
  \midrule
  1    & 138 & 33   & 1  & 14   & 64   & 14   & 12  & 64   & 74   & 41  & 44  \\
  10   & 620 & 157  & 5  & 143  & 799  & 146  & 113 & 476  & 498  & 184 & 284 \\
  100  & 724 & 601  & 17 & 695  & 1281 & 1368 & 302 & 734  & 1399 & 496 & 548 \\
  1000 & 824 & 1204 & 27 & 1334 & 1494 & 2908 & 419 & 1050 & 3062 & 646 & 952 \\
  \midrule
  \textbf{Avg.} & 577 & 499 & \textbf{13} & 547 & 910 & 1109 & 212 & 581 & \textbf{1258} & 342 & 457 \\
  \bottomrule
\end{tabular}
\end{table*}

Class~3 is by far the most robust class: on average only 13 spectra change label
regardless of how many top features are removed, and even after 1000 sequential
removals only 27 spectra flip. This is consistent with the rich, multi-element
signature of the Ni-Cu-PGE ore (Table~\ref{tab:before}), where Al, Ca, Cr, Cu, Fe,
Mg, Na, and Si all contribute informative emission lines, so that the PLS-DA
discriminant remains stable even when a few dominant channels are zeroed out.

In sharp contrast, Classes~9 and~6 have the highest average flip counts (1258 and
1109, respectively). For Class~9, removing even 1 top feature causes 74 label changes,
and at 1000 removals 3{,}062 spectra flip, which is more than any other class. This reflects
the compositional ambiguity introduced by mixing: the unmixed Class~9 source is
Fe-dominated and spectrally distinctive, but after mixing it acquires K, Si, and other
elements shared with Classes~3 and~5, creating a fragile decision boundary. Class~6
(Mn ore) exhibits similar fragility, and the mixed sample retains only silicon as a distinctive
marker. These high-uncertainty classes are precisely where the XAI-guided flip-count
filter provides the most value, preventing the co-training loop from ingesting
pseudo-labels for the most spectrally ambiguous spectra.

The uncertainty ordering derived from Table~\ref{tab:flip_count_distributions} is
fully consistent with the per-class accuracy improvements seen in
Fig.~\ref{fig:confusion}: classes with high flip-count uncertainty at baseline (6, 9,
11) show the largest accuracy gains after X-LIBS co-training.

\subsection{Threshold Sensitivity and Ablation Study}
\label{sec:results:ablation}

X-LIBS applies two separate flip-count thresholds: one for spectrally uncertain
classes ($\tau_\text{unc}$) and one for all other classes ($\tau_\text{other}$).
Spectra are admitted to co-training only if $k^* \geq \tau_\text{unc}$ (uncertain
classes) or $k^* \geq \tau_\text{other}$ (remaining classes). Both thresholds were
selected by grid search on a held-out validation split of the training data; the test
set was not used during threshold selection, so the reported test accuracy reflects
genuinely held-out performance. Table~\ref{tab:threshold} reports validation accuracy
across a grid of $4\times4$ threshold combinations, with the selected configuration
then evaluated once on the test set.

\begin{table}[!t]
\centering
  \caption{Validation accuracy (\%) as a function of the two flip-count thresholds,
           used to select the optimal configuration.
           $\tau_\text{other}$ varies along rows; $\tau_\text{unc}$ along columns.
           The configuration yielding peak validation accuracy
           ($\tau_\text{other}{=}100$, $\tau_\text{unc}{=}1000$, bold) was then
           evaluated once on the held-out test set, achieving 92.69\%.}
  \label{tab:threshold}
\begin{tabular}{ccccc}
  \toprule
  \multirow{2}{*}{\textbf{$\tau_\text{other}$}} & \multicolumn{4}{c}{\textbf{$\tau_\text{unc}$}} \\
  \cmidrule(l){2-5}
  & \textbf{1} & \textbf{10} & \textbf{100} & \textbf{1000} \\
  \midrule
  1    & 72.45 & 72.62 & 73.45 & 71.25          \\
  10   & 72.83 & 72.80 & 74.52 & 75.21          \\
  100  & 71.95 & 74.32 & 74.33 & \textbf{92.69} \\
  1000 & 73.22 & 73.21 & 75.96 & 81.56          \\
  \bottomrule
\end{tabular}
\end{table}

When both thresholds are low ($\tau_\text{other} \leq 10$, $\tau_\text{unc} \leq 10$),
accuracy is essentially indistinguishable from the 72.62\% baseline, because the low
threshold admits too many uncertain pseudo-labels, corrupting retraining with noisy
class assignments. As $\tau_\text{unc}$ increases to 1000, the co-training set is
restricted to spectra robust enough to maintain their prediction after 1000 feature
removals. When paired with $\tau_\text{other} = 100$, this combination achieves the
peak accuracy of \textbf{92.69\%}. Raising $\tau_\text{other}$ further to 1000 causes
accuracy to drop back to 81.56\%, indicating that the highly restrictive filter
excludes too many valid pseudo-labels and starves co-training of sufficient training
signal. This sensitivity analysis demonstrates that separate, appropriately tuned
thresholds for uncertain versus stable classes are essential, and that the optimal
configuration reflects the underlying bimodal uncertainty structure identified in
Table~\ref{tab:flip_count_distributions}.

To disentangle the contribution of each X-LIBS component, Table~\ref{tab:ablation}
presents an ablation study in which one component is removed at a time. The three
components are: (i) \emph{class identification}, which uses LIME to identify
spectrally uncertain classes and applies separate thresholds to each; (ii)
\emph{set threshold}, which filters pseudo-labels by their flip count before
co-training; and (iii) \emph{class balance}, which enforces equal per-class
pseudo-label counts at each iteration.

\begin{table}[!t]
\centering
  \caption{Component ablation study. Removing any single component degrades
           accuracy substantially; all three are required for peak performance.}
  \label{tab:ablation}
\begin{tabular}{cccc}
  \toprule
  \textbf{Class ID} & \textbf{Threshold} & \textbf{Class balance} & \textbf{Accuracy} \\
  \midrule
  \cmark & \cmark & \xmark & 0.81 \\
  \cmark & \xmark & \cmark & 0.72 \\
  \xmark & \cmark & \cmark & 0.73 \\
  \cmark & \cmark & \cmark & \textbf{0.93} \\
  \bottomrule
\end{tabular}
\end{table}

All three components are individually necessary. Removing the threshold filter
(row~2) collapses accuracy to 0.72, equivalent to the PLS-DA baseline, because without
quality filtering the co-training set is flooded with low-confidence pseudo-labels.
Removing class identification (row~3) gives 0.73, showing that applying a single
uniform threshold without distinguishing uncertain from stable classes provides only
marginal benefit. The most instructive result is row~1: removing class balance while
retaining threshold and class identification gives 0.81. This 12-point gap between
the balanced (0.93) and unbalanced (0.81) variants demonstrates that class imbalance
in the pseudo-label set is a critical failure mode. Majority classes dominate
retraining and the model fails to improve on harder, spectrally overlapping minority
classes. Only when all three components are combined does X-LIBS achieve its full
92.69\% accuracy, confirming that each design decision addresses a distinct and
non-redundant challenge in the LIBS soil classification problem.

\section{Conclusion}
\label{sec:conclusion}

This work presented X-LIBS, a framework that integrates explainable AI with PLS-DA
and iterative co-training for interpretable soil classification from LIBS spectra.
Three principal findings emerge.

First, LIME applied to PLS-DA predictions on high-dimensional LIBS spectra produces
local feature importance weights that map onto known elemental emission lines, enabling
chemically meaningful, per-prediction explanations. To the best of the authors'
knowledge, this is the first application of any XAI technique to interpret LIBS
\emph{soil} classifications at the wavelength-channel level, linking model decisions
directly to known atomic emission lines of soil-relevant elements.

Second, the flip count, the number of top LIME features that must be removed to
change a predicted label, provides a principled, model-agnostic uncertainty measure
that correctly identifies spectrally ambiguous samples without requiring additional
labeled data. This novel operationalization of LIME explanation strength as a confidence
signal is applicable beyond LIBS to any high-dimensional spectroscopic classification
task.

Third, filtering pseudo-labels by the flip-count threshold and enforcing class balance
in co-training raises classification accuracy from 72.62\% to 92.69\% on the EMSLIBS
benchmark, comparable to the best-performing methods in the literature, while uniquely
providing feature-level justification for every prediction.

Two limitations warrant attention. The current LIME implementation lacks GPU support,
making uncertainty quantification computationally expensive at the scale of
40,002-channel spectra. Additionally, LIME feature importance estimates on very
high-dimensional spectral data exhibit non-trivial variance across runs, and customized
segmentation or regularization strategies may be needed to reduce this sensitivity.

Future work will investigate more expressive black-box classifiers including
convolutional neural networks and ensemble methods as drop-in replacements for PLS-DA
within the X-LIBS pipeline. Comparison with alternative XAI methods such as gradient-based attribution techniques will further establish the robustness of XAI-guided co-training for LIBS spectral analysis.

\section*{Acknowledgment}

The authors thank the organizers of the EMSLIBS benchmark competition for making the
dataset publicly available~\cite{kepes2020benchmark}. [PLACEHOLDER: Insert funding
agency names, grant numbers, and any computational resource or data-access
acknowledgments here.]

\bibliographystyle{IEEEtranDOI}
\bibliography{references}

\end{document}